\documentclass[12pt]{iopart}

\usepackage{iopams}
\expandafter\let\csname equation*\endcsname\relax
\expandafter\let\csname endequation*\endcsname\relax
\usepackage{amsmath}
\usepackage{graphicx}
\usepackage{xcolor}

\DeclareMathOperator{\erfc}{erfc}

\renewcommand{\vec}{\mathbf}

\begin{document}

\title[Entropic contribution to phenotype fitness]{Entropic contribution to phenotype fitness}

 \author{Pablo Catal\'an$^{1,2}$, Juan Antonio Garc\'{\i}a-Mart\'{\i}n$^{2,3}$, Jacobo Aguirre$^{2,4}$, Jos\'e A. Cuesta$^{1,2,5}$, Susanna Manrubia$^{2,6}$}

\address{$^1$ Departamento de Matem\'aticas, Universidad Carlos III de Madrid, Madrid, Spain}
\address{$^2$ Grupo Interdisciplinar de Sistemas Complejos (GISC), Madrid, Spain}
\address{$^3$ Bioinform\'atica para Gen\'omica y Prote\'omica. Centro Nacional de Biotecnolog\'{\i}a (CNB-CSIC), Madrid, Spain}
\address{$^4$ Centro de Astrobiolog\'{\i}a (CAB), CSIC-INTA, Ctra. de Ajalvir km 4, Torrej\'on de Ardoz, Madrid, Spain}
\address{$^5$ Instituto de Biocomputaci\'on y F\'{\i}sica de Sistemas Complejos (BIFI), Universidad de Zaragoza, Zaragoza, Spain}
\address{$^6$ Departamento de Biolog\'{\i}a de Sistemas. Centro Nacional de Biotecnolog\'{\i}a (CNB-CSIC), Madrid, Spain}
\ead{smanrubia@cnb.csic.es}
\vspace{10pt}



\begin{abstract}
All possible phenotypes are not equally accessible to evolving populations. In fact, only phenotypes of large size, i.e. those resulting from many different genotypes, are found in populations of sequences, presumably because they are easier to discover and maintain. Genotypes that map to these phenotypes usually form mostly connected genotype networks that percolate the space of sequences, thus guaranteeing access to a large set of alternative phenotypes. Within a given environment, where specific phenotypic traits become relevant for adaptation, the replicative ability of a phenotype and its overall fitness (in competition experiments with alternative phenotypes) can be estimated. Two primary questions arise: how do phenotype size, reproductive capability and topology of the genotype network affect the fitness of a phenotype? And, assuming that evolution is only able to access large phenotypes, what is the range of unattainable fitness values? 
In order to address these questions, we quantify the adaptive advantage of phenotypes of varying size and spectral radius in a two-peak landscape. We derive analytical relationships between the three variables (size, topology, and replicative ability) which are then tested through analysis of genotype-phenotype maps and simulations of population dynamics on such maps. Finally, we analytically show that the fraction of attainable phenotypes decreases with the length of the genotype, though its absolute number increases. The fact that most phenotypes are not visible to evolution very likely forbids the attainment of the highest peak in the landscape. Nevertheless, our results indicate that the relative fitness loss due to this limited accessibility is largely inconsequential for adaptation.
\end{abstract}

%
\vspace{2pc}

%
\submitto{\jpa}
%
%
%

\section{Introduction}

Our understanding of how genotypes map onto phenotypes, functional pieces and, eventually, whole organisms, has been boosted by studies of simple genotype-to-phenotype (GP) maps (reviews in \cite{stadler:2006,ahnert:2017,manrubia:2021}). At odds with a pre-sequencing era view where the mapping between sequence and function was thought to be one-to-one \cite{ogbunugafor:2020}, biologically relevant GP maps are many-to-many, with a huge redundancy that has been deeply explored to show, in particular, that a specific phenotype can be achieved from an astronomically large number of genotypes \cite{louis:2016}. 

The set of genotypes that map to a specific phenotype typically form large networks where genotypes are nodes and links represent a mutational move \cite{bastolla:2003,ciliberti:2007,matias-rodrigues:2011,schultes:2000,aguirre:2011}. Assigning a  unique phenotype to each genotype partitions the space of genotypes into a set of non-overlapping, but linked, phenotypes, and induces a network-of-networks organization in genotype spaces \cite{yubero:2017}. Knowing the topological features of genotype networks \cite{aguirre:2011,ahnert:2017} is essential to perform an accurate description of evolutionary dynamics \cite{aguirre:2018}. Phenotype size, defined as the number of genotypes that map onto that phenotype, follows a very skewed distribution, with a small fraction of the largest phenotypes covering most of genotype space; in many cases, the distribution of phenotype sizes is well fit by a lognormal function \cite{dingle:2015,cuesta:2017,catalan:2018,garcia-martin:2018,villanueva:2022}. Both in numerical and empirical studies \cite{jorg:2008,dingle:2015,villanueva:2022}, observed phenotypes are typically large, while most phenotypes are never visited through blind evolutionary searches. A network representation of related genotypes, instead of a phylogenetic tree, can bring out the existence of cycles that reveal parallel or convergent evolution \cite{wagner:2014PRSB}. Also, the degree distribution of genotype networks can be put in direct correspondence with the robustness of a phenotype: the higher the average degree, the lower the effect of mutations, on average. A remarkable feature identified in multiple GP maps is a linear correlation between phenotype robustness (or average degree of the genotype network) and the logarithm of phenotype size \cite{aguirre:2011,greenbury:2016,greenbury:2014,catalan:2017,dallolio:2014}. 

Despite its evolutionary relevance, the adaptive effects of phenotype size remain largely unexplored from a formal viewpoint. When we think of ``fitness'' of a population, more often than not we recreate the classical fitness landscape that Wright introduced almost a century ago \cite{wright:1932}. In this widespread metaphorical representation, devised long before the community became acquainted with the structure and organization of molecular populations, fitness optima corresponded to hilltops in a two-dimensional landscape: adaptation was a parsimonious process that proceeded always uphill and, once mutation-selection equilibrium was attained, populations were forever sitting at the top of the hill. Though this representation cannot, by construction, include the adaptive effects of robustness in phenotype fitness \cite{catalan:2017} or environmental variation \cite{mustonen:2009}, Wright's fitness landscapes still condition most expectations on the outcome of the evolutionary process \cite{laland:2014,svensson:2012,aguirre:2022}. 

The previous criticism notwithstanding, the last two decades have witnessed an increase in the number of works dealing with the effect of phenotype size in adaptation; terms such as entropy, phenotypic redundancy or landscape flatness have been used as synonyms of size. A pioneering work by Schuster and Swetina \cite{schuster:1988} discussed cases of competition between two phenotypes where the sequence with the highest selective value had a less efficient neighborhood than that with the second largest selective value; they demonstrated that too low a robustness could be fatal at high mutation rates. In a study of spatial gene regulation during development, it was shown that the convergence of finite populations to the maximally fit phenotype was compromised by the multiplicity or entropy of solutions \cite{khatri:2009}. The survival of the flattest was also considered a surprising effect where a population of replicators would select regions of the landscape of lower fitness but ``flatter'', at sufficiently high mutation rates \cite{wilke:2001Nat}. In a related work where this effect was empirically tested with viroids, the authors stated that fitness should not always be associated with fast replication, and that fitness can indeed be maximized by reducing the impact of mutations on a phenotype \cite{codoner:2006}. Computational analyses of population dynamics with mutation on the genotype and selection on the phenotype have further clarified the relevance of phenotype size, in phenomena termed the ascent of the abundant \cite{cowperthwaite:2007} or the arrival of the frequent \cite{schaper:2014,catalan:2020}. 

There is thus broad evidence that evolving populations do tend towards an optimum that is (at least) a combination of replicative ability and phenotype redundancy. In this work, we quantitatively derive the contribution of both terms to the overall fitness under simple conditions. We begin by presenting numerical and theoretical evidence of some important properties of genotype networks, and formally study the case of a population evolving on a network formed by two phenotypes of different size and replicative ability, much in the spirit of \cite{schuster:1988}. Our aim is to establish the conditions under which the population would transition from one phenotype to the other, and express the transition point as a function of phenotype properties. In order to provide a numerical illustration of the theoretical results, we explore two GP maps of different complexity. First, we use the RNA sequence-to-secondary structure (S3) map, a paradigmatic example~\cite{fontana:1993,ancel:2000,schuster:2006} for which precise numerical and theoretical results regarding the topological nature of its phenotype networks are available. Second, we revisit a pattern-generating version of toyLIFE~\cite{catalan:2020} that we call toyLIFE T2P. toyLIFE is a multilevel GP map that relies on the simple hydrophobic-polar (HP) model for basic interactions~\cite{arias:2014,catalan:2017tesis,catalan:2018} and that, despite its complexity, displays qualitative properties analogous to RNA. We close by analyzing and discussing the implications that selection for large phenotypes has in the attainment of phenotypes of sufficiently high replicative ability.   

\section{Genotype networks}

Here, a phenotype is defined as a connected network of genotypes with the same replicative ability. Genotypes are sequences of letters taken from a given alphabet. Two nodes are linked if they differ in one position of their sequences. The number of neighbors of a given node, its degree $k_i$, is a measure of robustness: the degree is low when point mutations tend to modify the phenotype of sequences one mutation away, while it is high for highly neutral sequences, whose phenotype is typically maintained under mutations. The average degree of a phenotype is defined as $\langle k \rangle = N^{-1}\sum_{i=1}^N k_i$, where $N$ is the phenotype size, or the number of nodes in its genotype network ${\bf G}$; ${\bf G}$ is the (symmetric) adjacency matrix of a connected (undirected) graph, whose elements are $G_{ij} = 1$ if nodes $i$ and $j$ are connected and $G_{ij} = 0$ otherwise, and whose topology is characterized by a spectral radius $\gamma$, the largest eigenvalue of $\bf G$.  Finally, we assume a constant environment; otherwise, both the precise set of nodes forming the network and/or the value of the replicative ability could change. 

\subsection{Evolutionary dynamics of replicator populations}

\begin{figure}[b!]
    \centering
    \includegraphics[width=130mm]{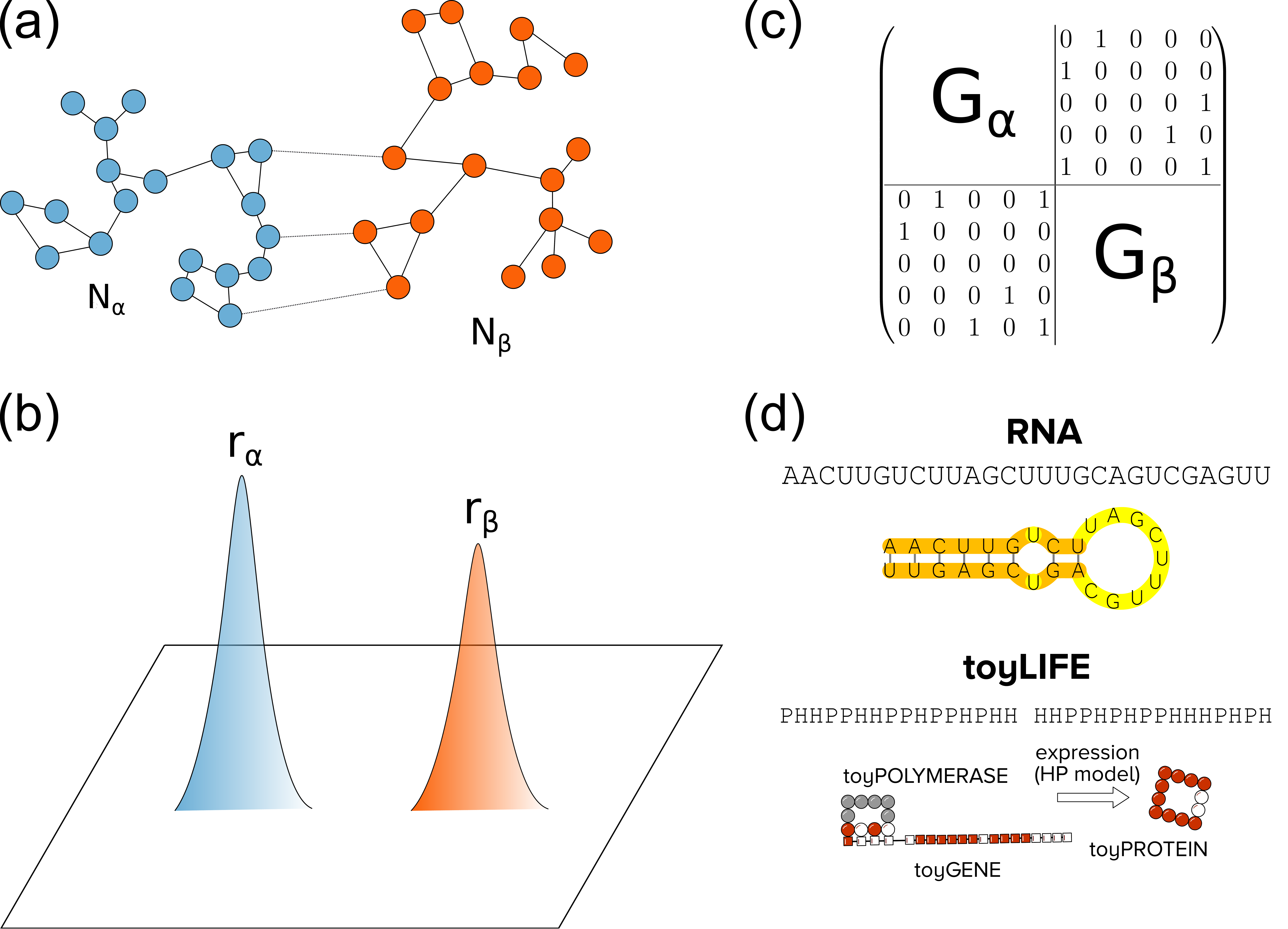}
    \caption{Schematic representation of the scenario considered in this work. A phenotype is characterized by a neutral network of genotypes, all of which share the same replicative ability $r$. (a) We consider a situation with two phenotypes, $\alpha$ and $\beta$, with networks of sizes $N_{\alpha}$ and $N_{\beta}$ and known topology, on (b) a two-peak landscape representing replicative abilities. (c) The joint adjacency matrix of the $\alpha+\beta$ system is formed by two diagonal blocks, each containing the adjacency matrix ${\bf G}_{\alpha}$, ${\bf G}_{\beta}$ of each phenotype (with spectral radii $\gamma_{\alpha}$ and $\gamma_{\beta}$) plus a number of off-diagonal terms that represent mutational connections between the two phenotypes. (d) The phenotypes of two GP maps will be computationally studied, RNA S3 and toyLIFE T2P. See main text for further details. 
    }
    \label{fig:cartoon}
\end{figure}

The evolution of a population of replicators on a fitness landscape, assuming discrete generations for simplicity, can be written as 
\begin{equation}
  \vec{n}(t)={\bf M} \vec{n}(t-1) = {\bf M}^t \vec{n}(0) = \sum_{i=1}^N
\lambda_i^t  (\vec{n} (0) \cdot \vec{u}_i) \vec{u}_i \, ,
\end{equation}
where $\vec{u}_i$ and $\lambda_i$ are the eigenvectors and eigenvalues of the (symmetric)
evolution matrix {\bf M}, and $\vec{n}(t)$ has length $N$ \cite{aguirre:2009,aguirre:2018}; $\vec{n}(0)$ is the initial condition.  
By definition, the nonnegative matrix {\bf M} is primitive (see below), so the Perron-Frobenius theorem ensures that, over time, the system evolves towards an asymptotic state characterised by the unique first (in decreasing order of eigenvalues) eigenvector $\vec{u}_1$. In biological terms, this state corresponds to the mutation-selection equilibrium. The components of $\vec{u}_1$ are all strictly positive and proportional to the asymptotic fraction of the total population at each node, while its associated eigenvalue $\lambda_1$ represents the asymptotic growth rate of the population. 

In a population of replicators that mutate with probability $0 < \mu < 1$ per genotype and replication cycle, matrix {\bf M} can be decomposed as

\begin{equation}
  {\bf M} = (1 - \mu) {\bf R} + \frac{\mu}{S} {\bf G R} \, ,
\label{eq:primitive}
\end{equation}
where ${\bf R}$ is the diagonal matrix $R_{ij} = r_i \delta_{ij}$, $r_i$ being the replicative ability of node (genotype) $i$. For a fixed phenotype, we will consider in this contribution that $r_i \equiv r$ for all $i$, where $r$ can be interpreted as the average number of copies of a given sequence in the next generation (time step). $S$ stands for the maximum number of neighbours of a genotype. When replicators are sequences of length $L$ whose elements are taken from an alphabet of $A \ge 2$ letters, the size of the genotype space is $m=A^L$, and $S=L(A-1)$. 

Matrices such as ${\bf M}$ in equation (\ref{eq:primitive}) are guaranteed to be
primitive if the network ${\bf G}$ is connected and the diagonal of ${\bf R}$ is strictly positive. Both conditions are fulfilled, by definition and because $r_i>0$ represent replicative values. 

Matrices $\bf M$ and $\bf G$ share eigenvectors (because ${\bf R}=r{\bf I}$), and their respective eigenvalues $\lambda$ and $\gamma$ are related through

\begin{equation}
\lambda = r \left[ (1-\mu) + \frac{\gamma \mu}{(A-1) L} \right] \, \, .
\label{eq:lg}
\end{equation}

\subsection{Competition between phenotypes}

Consider two phenotypes $\alpha$ and $\beta$, each represented by a different network, with parameters $N_{\alpha/\beta}$, $r_{\alpha/\beta}$ and $\gamma_{\alpha/\beta}$. Further assume that the two phenotypes are mutually accessible through single mutations (excluding deletions and insertions) from one or a few nodes in their networks (see figure \ref{fig:cartoon}). The matrix $\bf M$ describing the evolution of a population of replicators in the two-peak landscape formed by the two phenotypes has a diagonal term for replication plus a topological contribution: two blocks along the diagonal, each corresponding to one of the phenotypes, and one or a few non-zero elements off the blocks representing the connections between nodes in different phenotypes just one mutation away (figure \ref{fig:cartoon}(c)). 

The question of which of the two phenotypes would be preferred by the population, and the point where most of the population would transition from one phenotype to the other was addressed in a similar scenario by Schuster and Swetina \cite{schuster:1988}. As a first approximation, let us assume, following those authors, that the two blocks are decoupled. In this case, the phenotype with the largest eigenvalue will be the preferred choice. That is, if a population occupies phenotype $\alpha$, it will transition to phenotype $\beta$ when $\lambda_{\beta} > \lambda_{\alpha}$: the eigenvalue $\lambda$ can be interpreted as the fitness of a phenotype. Note that, in this interpretation, fitness $\lambda$ results from a non-trivial combination of the topological properties in Fig. \ref{fig:cartoon}a and the replication rate, as represented in Fig. \ref{fig:cartoon}b. Indeed, recalling Eq.~\eqref{eq:lg}, we obtain a relationship between the replication rate and the spectral radii of the two phenotypes, stating that the transition $\alpha \to \beta$ will occur if
\begin{equation}
\frac{r_{\beta}}{r_{\alpha}} > \frac{1-\mu + \gamma_{\alpha} \mu (A-1)^{-1} L^{-1}}{1-\mu + \gamma_{\beta} \mu (A-1)^{-1} L^{-1}} \simeq
1+\frac{\mu}{(A-1)L} (\gamma_{\alpha}-\gamma_{\beta}) \, ,
\label{eq:r-gamma}
\end{equation}
the last approximation holding for $\mu \ll 1$. 

This inequality will more accurately describe the transition the less connected the two phenotypes are. As the number of connections between phenotypes increases, the separation in two blocks of the adjacency matrix becomes progressively blurred and the description in terms of the two corresponding eigenvalues worsens. This issue will be further discussed and numerically tested in forthcoming sections. 

\subsection{Bounds to the spectral radius}

The spectral radius is a measure of a network's topology. Its value admits various bounds as a function, in particular, of the average $\langle k \rangle$, maximum $k_{\rm max}$ and minimum $k_{\rm min}$ degree in the network~\cite{nimwegen:1999},

\begin{equation}
k_{\rm{min}} \le \langle k \rangle \le \gamma \le  k_{\rm{max}} \, ,
\label{eq:gbounds}
\end{equation}
equalities holding for homogeneous networks, where all nodes have the same degree. 

\subsubsection{Spectral radius and the mean degree of a graph.}

The bound $\langle k \rangle \le \gamma$ is a known result that can be easily proven. The spectral radius of a real symmetric matrix ${\bf A}$ is defined as
\begin{equation}
    \rho({\bf A})=\max_{\|{\bf x}\|=1}{\bf x}^{\sf T}{\bf Ax}.
\end{equation}
Thus, if $\bf A$ satisfies the conditions of Perron-Frobenius's theorem (i.e., if the underlying graph is connected and not multipartite) and $\bf v$ is the (unique) eigenvector associated to $\rho({\bf A})$ (the largest eigenvalue), then
\begin{equation}
    {\bf x}^{\sf T}{\bf Ax}<{\bf v}^{\sf T}{\bf Av}=\rho({\bf A}), \qquad {\bf x}\ne{\bf v}.
\end{equation}
So, if $\bf G$ is the adjacency matrix of a connected graph and $\gamma$ is its spectral radius, and if we take for ${\bf x}$ a uniform vector with components $N^{-1/2}$, then
\begin{equation}
    {\bf x}^{\sf T}{\bf Gx}=\langle k\rangle<\gamma
    \label{eq:kbound}
\end{equation}
as long as ${\bf x}\ne{\bf v}$ (i.e.~the graph is not regular).

We can improve this lower bound for $\gamma$ by repeating the argument with ${\bf G}^2$ rather that ${\bf G}$. As ${\bf x}^{\sf T}{\bf G}^2{\bf x}=\|{\bf Gx}\|^2$ and $({\bf Gx})_i=k_iN^{-1/2}$, then we obtain $\langle k^2\rangle<\gamma^2$ for any nonregular graph. If $\sigma^2$ denotes the variance of the degree distribution, then
\begin{equation*}
    \langle k^2\rangle=\langle k\rangle^2+\sigma^2=\langle k\rangle^2\left(1+\frac{\sigma^2}{\langle k\rangle^2}\right),
\end{equation*}
and the lower bound becomes
\begin{equation}
    \gamma>\langle k\rangle\sqrt{1+\frac{\sigma^2}{\langle k\rangle^2}}\ge
    \langle k\rangle+\frac{\sigma^2}{2\langle k\rangle^2}
    \label{eq:sigmabound}
\end{equation}
(the last inequality follows from the inequality $\sqrt{1+x}\ge 1+(x/2)$, valid for all $x\ge 0$). Inequalities \eqref{eq:kbound} and \eqref{eq:sigmabound} yield important relationships between the average degree of a graph and its spectral radius, which determines the asymptotic state of a population of replicators, as described above. For homogeneous graphs, where $k_i=k=\langle k \rangle$ for all $i$, $\langle k \rangle = \gamma$ holds. The question is, how far from homogeneous are genotype networks? Are the previous relationships relevant to predict the evolutionary behavior of a population on these networks? 

\subsubsection{Average degree and network size.} Genotype-to-phenotype (GP) maps have been broadly used to generate genotype networks and to characterize the topological properties the map confers to sequence spaces~\cite{stadler:2006,wagner:2011,ahnert:2017,manrubia:2021}. Some of the quantities derived, most often numerically, seem to be quasi-universal, in the sense that they are repeatedly found in a variety of GP maps. Such is the relationship between the average degree of a genotype network and its size, which is largely independent of the specific definition of phenotype: $\langle k \rangle \sim \log N$.

This relationship can be heuristically calculated taking as example the case of the RNA sequence-to-secondary structure (S3) map~\cite{aguirre:2011}, though the results are more general. First, we recall that an excellent estimation of the number $N$ of genotypes folding into a given RNA secondary structure can be obtained by calculating the so-called versatility of each position along the sequence. The versatility $v_j$ of site $j$, $j=1, \dots, L$, is defined as the number of mutations (out of the total size $A$ of the alphabet) that site $j$ accepts, averaged over all sequences in the network~\cite{manrubia:2017,garcia-martin:2018,martin:2022}, from which the size of the phenotype is estimated as
\begin{equation}
    N = \prod_{j=1}^L v_j.
    \label{eq:Nversatilities}
\end{equation}
Numerical comparison between this estimation and the exhaustive enumeration of genotypes in a phenotype yields an excellent agreement~\cite{garcia-martin:2018,martin:2022}. Asymptotically, the size of RNA secondary structures admits a two-versatility approximation \cite{cuesta:2017,garcia-martin:2018} that distinguishes just two different structural elements, paired and unpaired nucleotides, each class admitting on average a number $v_p$ and $v_u$ of neutral mutations, respectively (see also \cite{huynen:1996b,reidys:2001}). In a previous contribution, it was shown that $\langle k \rangle \propto \log N$ using this approximation. 

\begin{figure}[b!]
    \centering
    \includegraphics[width=150mm]{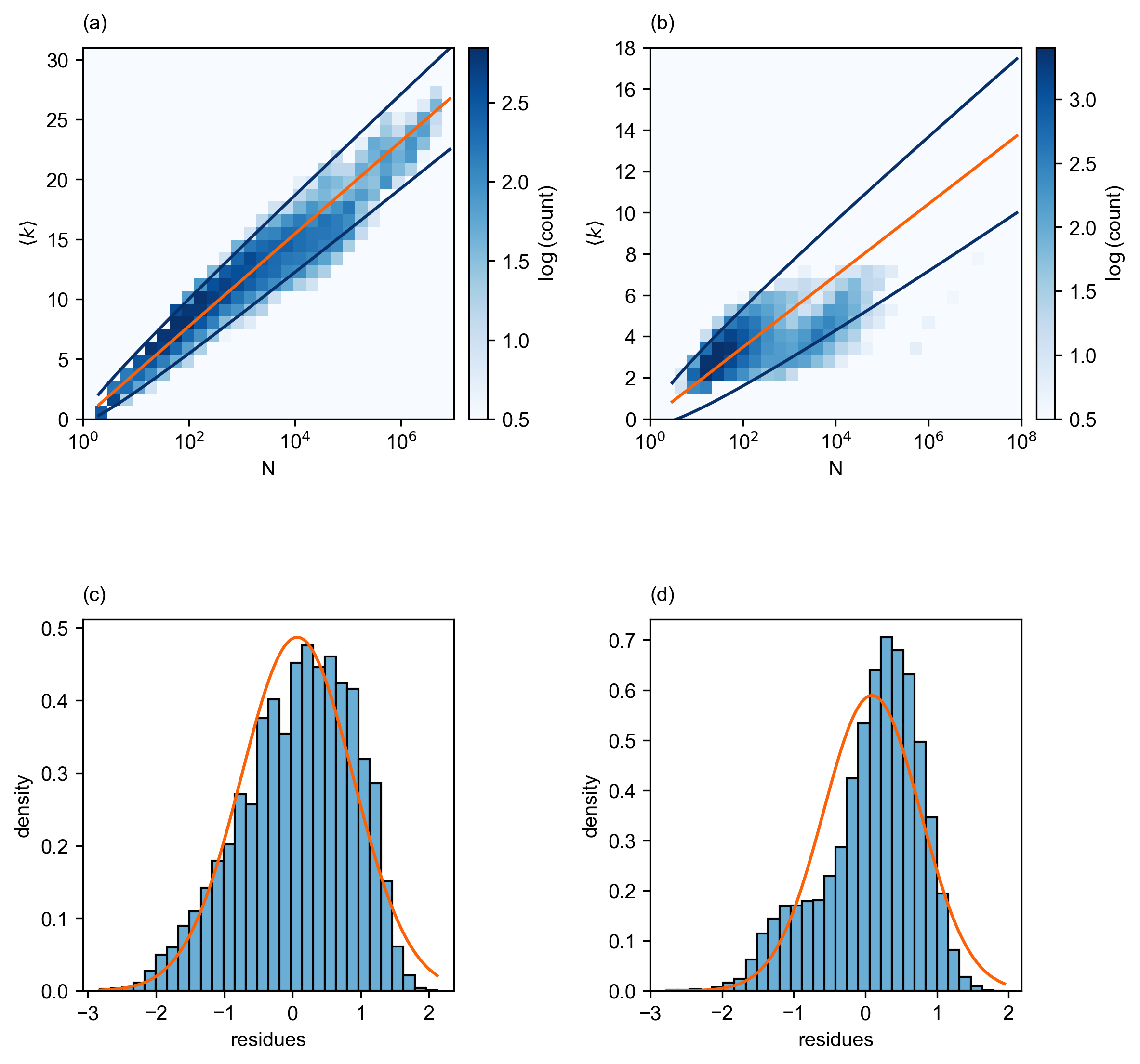}
    \caption{Relationship between $\langle k \rangle$ and $\log N$, in (a) RNA S3 $L=16$ and  (b) toyLIFE T2P. The orange line represents the average of $\langle k \rangle$ as given by equation \eqref{eq:RNA-kN}, $\nu_{\langle k \rangle}=c \log N$, while the dark blue lines represent the $95\%$ confidence interval, $\nu_{\langle k \rangle} \pm 1.96\sigma_{\langle k \rangle}$, with $\sigma_{\langle k \rangle}=\kappa(\log N)^{1/2}$. For RNA (a), $c=3.863,\kappa=0.822,$ (a) while, for toyLIFE, $c=1.7375, \kappa=0.684$ (b). Lower panels represent the distribution of residues of the least-squares fit $(\log N)^{-1/2}\langle k \rangle=c(\log N)^{1/2}$ for (c) RNA and (d) toyLIFE. Failure to fit to a Gaussian distribution might arise from the small length of genotypes or, especially in toyLIFE, perhaps be a constitutive property of the model. 
    }
    \label{fig:kavgvslogN}
\end{figure}

Nevertheless, an argument justifying the dependence of the average degree on $\log N$ can be obtained directly from \eqref{eq:Nversatilities} if we interpret this expression as the product of $N$ random variables. By taking logarithms, $\log N$ can be regarded, in the limit $L \to \infty$, as a normal random variable $\log N\approx L\nu_{\log}+L^{1/2}\sigma_{\log}\xi$, where $\nu_{\log}\equiv\langle\log v\rangle$, $\sigma_{\log}\equiv\langle(\log v-\nu_{\log})^2\rangle$, and $\xi$ is a random variable distributed as $\xi\sim\mathcal{N}(0,1)$.
On the other hand,
\begin{equation}
\langle k \rangle = \sum_{j=1}^L(v_j -1),
\end{equation}
is another random variable $\langle k\rangle\approx L(\nu-1)+L^{1/2}\sigma\xi'$, where $\nu\equiv\langle v\rangle$, $\sigma\equiv\langle(v-\nu)^2\rangle$, and $\xi'\sim\mathcal{N}(0,1)$. Thus, combining both expressions,
\begin{equation}
\langle k \rangle \approx c \log N + \kappa(\log N)^{1/2}\eta, \qquad \eta\sim\mathcal{N}(0,1),
\label{eq:RNA-kN}
\end{equation}
where the coefficients $c$ and $\kappa$ depend on statistical properties of the distribution of versatilities in the GP map, but not on the particulars of a specific phenotype (that information is hidden in the random variable $\eta$). We can estimate the coefficients in the expression~\eqref{eq:RNA-kN} using maximum likelihood to fit a normal distribution to the empirical data of a set of $P$ phenotypes. The resulting formulas for them are
\begin{equation}
    c=\frac{\sum_{i=1}^P\langle k_i\rangle}{\sum_{i=1}^P\log N_i}, \qquad
    \kappa^2=\sum_{i=1}^P\frac{\big(\langle k_i\rangle-c\log N_i\big)^2}{P\log N_i}.
\end{equation}
Figure \ref{fig:kavgvslogN} shows the fit of equation \eqref{eq:RNA-kN} to RNA S3, $L=16$, and toyLIFE T2P, with very good results. These also extend previous numerical analysis of RNA sequences of length $L=12$, which showed that equation (\ref{eq:RNA-kN}) yields, to first order in $\log N$, an excellent fit to numerical results (see figure~3B in~\cite{aguirre:2011}).

In view of the success of the versatility model \eqref{eq:Nversatilities} in describing the size distribution of different GP maps \cite{garcia-martin:2018}, equation~\eqref{eq:RNA-kN} turns out to be more general than the above derivation, with the RNA model in mind, might suggest. Independent analyses have shown that the proportionality between average degree and phenotype size is not limited to RNA S3, as it has been numerically obtained in simple models of protein folding~\cite{greenbury:2016}, in a model for protein quaternary structure~\cite{greenbury:2014} and in toyLIFE~\cite{catalan:2017}. Interestingly, it also describes well some empirical observations, as the relationship observed in a genotype network reconstructed from short haplotypes in the human chromosome 22~\cite{dallolio:2014}. Altogether, these results strongly suggest that $\langle k \rangle \propto \log N$ may be a quasi-universal property of biologically realistic GP maps and fundamentally related to the distribution of phenotype sizes, as the derivation of both results in the framework of the versatility model strongly suggests. 

\section{Genotype network topology in numerical GP maps}
\label{sec:numerical-topology}

The theory derived in the previous section relates, on the one hand, the fitness $\lambda$ of a phenotype with its replicative ability and its spectral radius, Eq. (\ref{eq:lg}), and, on the other hand, the average degree $\langle k \rangle$ and the $\log$ size of a phenotype, Eq. (\ref{eq:RNA-kN}). Both expressions are further related through the inequality $\langle k \rangle \le \gamma$.

Numerical simulations in this Section are devoted to explore the topological properties of two representative GP maps of different complexity, RNA S3 and toyLIFE T2P. Our eventual aim is to check how close the average degree $\langle k \rangle $ is to the spectral radius $\gamma$ of a genotype network; should the approximation $\langle k \rangle \simeq \gamma$ be feasible, we could derive an approximate relationship between the fitness of a phenotype and its size. 

\subsection{RNA}

\begin{figure}[b!]
    \centering
    \includegraphics[width=150mm]{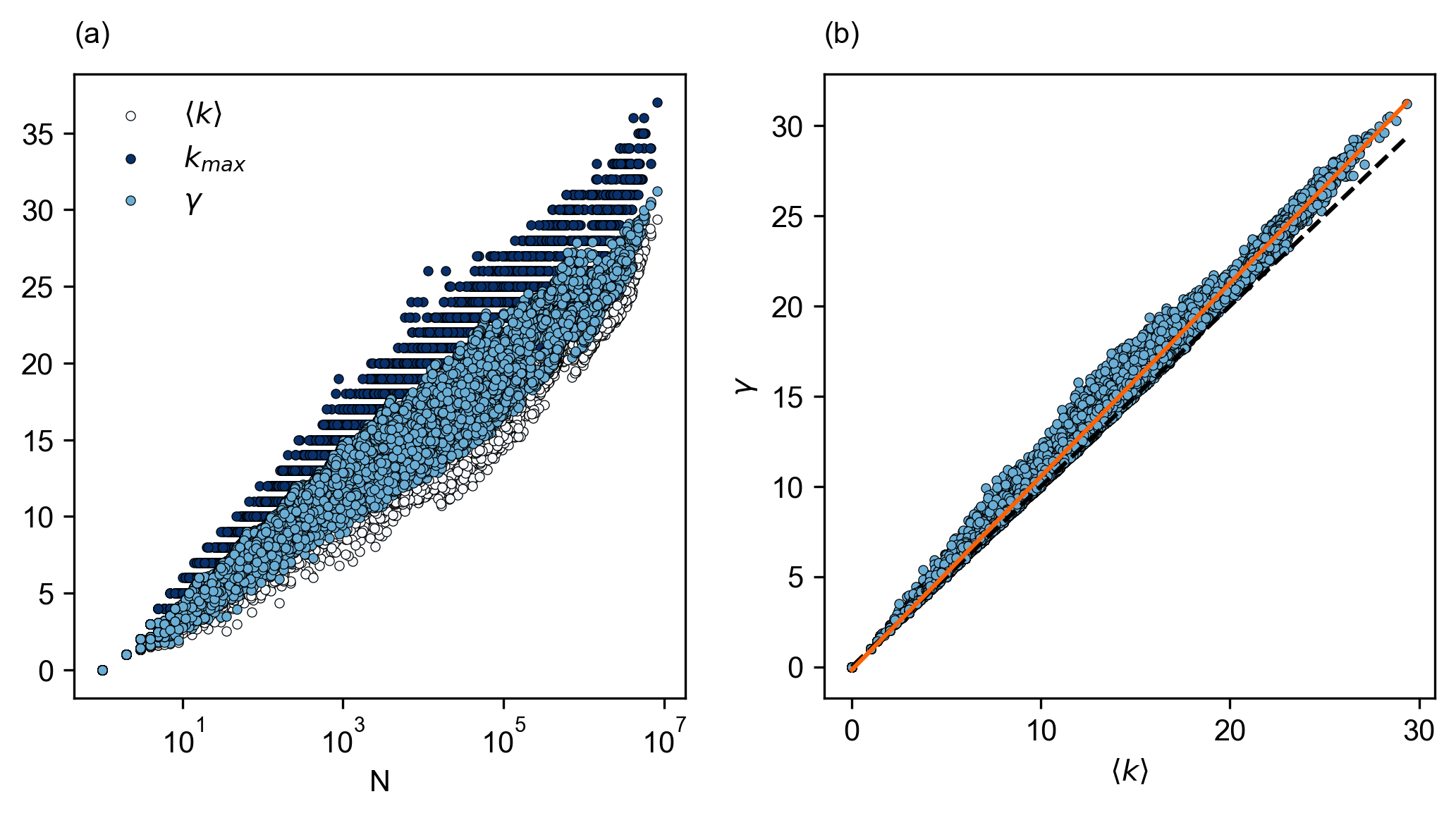}
    \caption{Topological quantities characterizing RNA S3 networks for $L=16$. (a) We represent the maximum and average network degree, as well as the network spectral radius, for connected components (CC) obtained through the exhaustive enumeration of the sequence space, as a function of CC size. (b) Connected component spectral radius $\gamma$ as a function of the corresponding average degree $\langle k \rangle$. The orange line represents a linear fit between the two measures: $\gamma = 1.07 \langle k \rangle-0.17$, with $R^2=0.99$. The dashed black line represents the line $\gamma=\langle k \rangle$, as a visual aid to confirm that $\gamma \ge \langle k \rangle$.
    }
    \label{fig:RNAtopology}
\end{figure}

We have exhaustively folded the space of RNA sequences of lengths $L=14, 15$, and $16$, mapped each sequence to its minimum-free-energy secondary structure, and separated each phenotype into connected components (CC) \cite{aguirre:2011,garcia-martin:2018}. The results obtained are comparable for the three genotype lengths  above (with 3,311, 8,792 and 23,091 CCs, respectively) and consistent with those obtained for $L=12$ \cite{aguirre:2011}. Each CC is a connected graph for which we have calculated the maximum degree $k_{\rm max}$, the average degree $\langle k \rangle$ of its nodes, and the spectral radius $\gamma$. Note that the maximum degree $k_{\rm max}$ corresponds to the node with the largest number of neutral neighbors in each CC and has to fulfill $k_{\rm max} \le (A-1) L$. The three quantities are jointly represented in Fig. \ref{fig:RNAtopology}(a) for $L=16$. In all cases, since all CC fulfill the conditions of the Perron-Frobenius theorem, the inequalities of equation (\ref{eq:gbounds}) hold. 

\begin{figure}[b!]
    \centering
    \includegraphics[width=150mm]{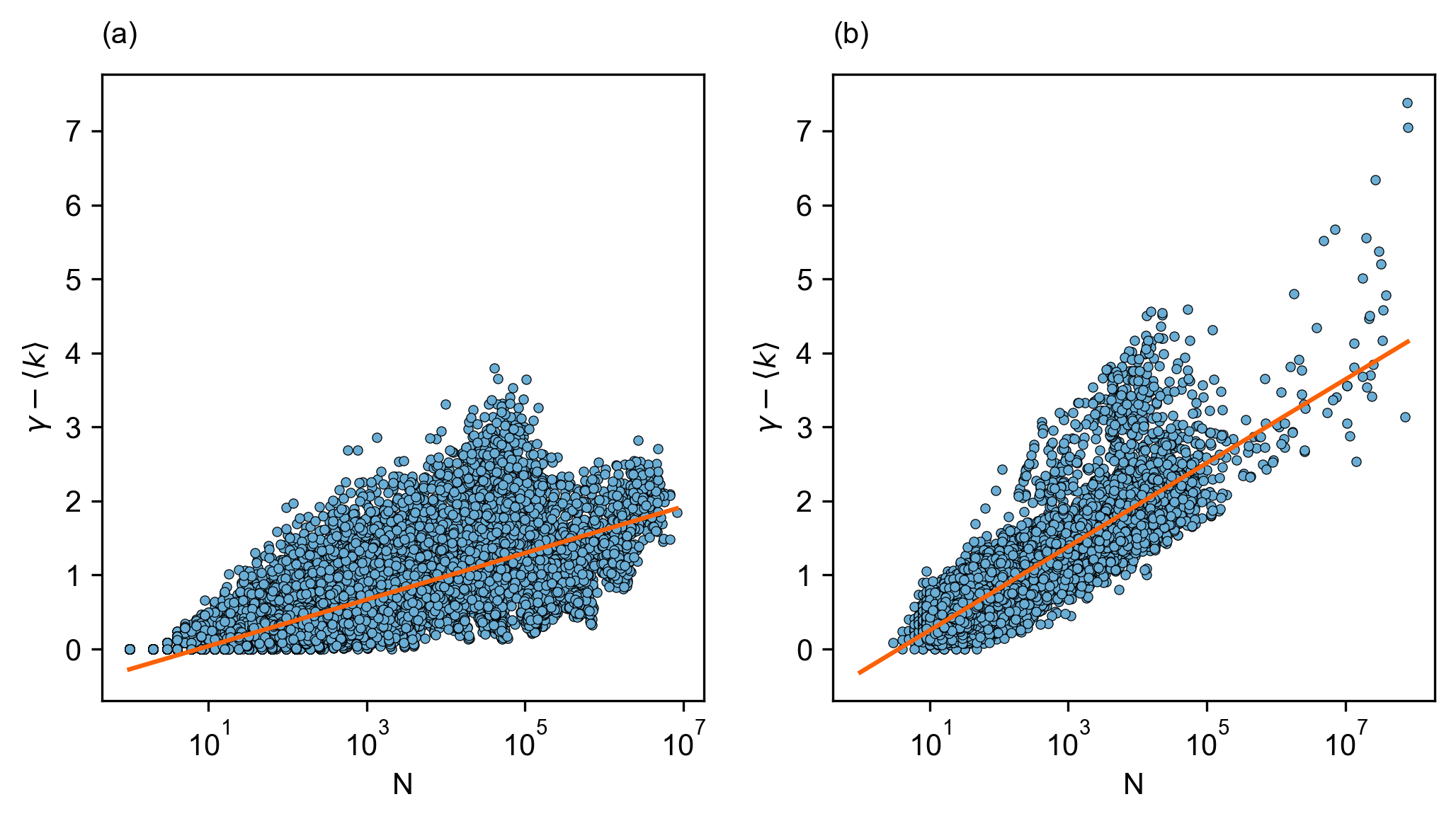}
    \caption{Difference between the spectral radius $\gamma$ and the average degree $\langle k \rangle$ as a function of network size $N$ for (a) RNA S3 with $L=16$ and (b) toyLIFE T2P. The lowest value $\gamma-\langle k \rangle=0$ corresponds to homogeneous networks, where all nodes have the same degree. In both cases, the orange line represents a linear fit: $\gamma-\langle k \rangle = 0.31 \log_{10} N -0.28$ for RNA S3 ($R^2=0.56$) and $\gamma-\langle k \rangle = 0.57 \log_{10} N-0.32$ for toyLIFE T2P ($R^2=0.76$).}
    \label{fig:RNAtopology-dif}
\end{figure}

Figure \ref{fig:RNAtopology}(b) depicts the calculated spectral radius as a function of the average degree. As it can be seen, both quantities are not only proportional, but also remain close for all values of $\langle k \rangle$ represented. Still, numerical data show a persistent dispersion due to specific (non-independent) phenotypic features that affect the degree distribution, such as size or the total number of paired nucleotides and their distribution within the considered RNA structure. Equation (\ref{eq:sigmabound}) made this dispersion explicit, giving a bound to the difference between both quantities, $\gamma - \langle k \rangle > \sigma^2/(2 \langle k \rangle)$. Figure \ref{fig:RNAtopology-dif}a illustrates the difference in the case of RNA, showing as well an increase with phenotype size $N$. 

In the limit $L \to \infty$, the distribution of structural elements in RNA secondary structures converges to a Gaussian distribution \cite{reidys:2002b,poznanovic:2014,cuesta:2017}. This fact does not eliminate the heterogeneity of the network for a fixed (typical) phenotype, but implies that the dispersion $\sigma$ is similar for different (typical) phenotypes. In this limit, since $c$ becomes independent of the phenotype, there is an additional approximation to the inequality in equation~(\ref{eq:r-gamma}) that can be performed. Substituting Eq.~(\ref{eq:RNA-kN}), we obtain

\begin{equation}
\frac{r_{\beta}}{r_{\alpha}} > 1+\frac{c \mu}{(A-1)L} \log\left(\frac{N_{\alpha}}{N_{\beta}}\right) + O(\sqrt{\log N})\, . 
\label{eq:r-N}
\end{equation}


\subsection{toyLIFE}

\begin{figure}[b!]
    \centering
    \includegraphics[width=150mm]{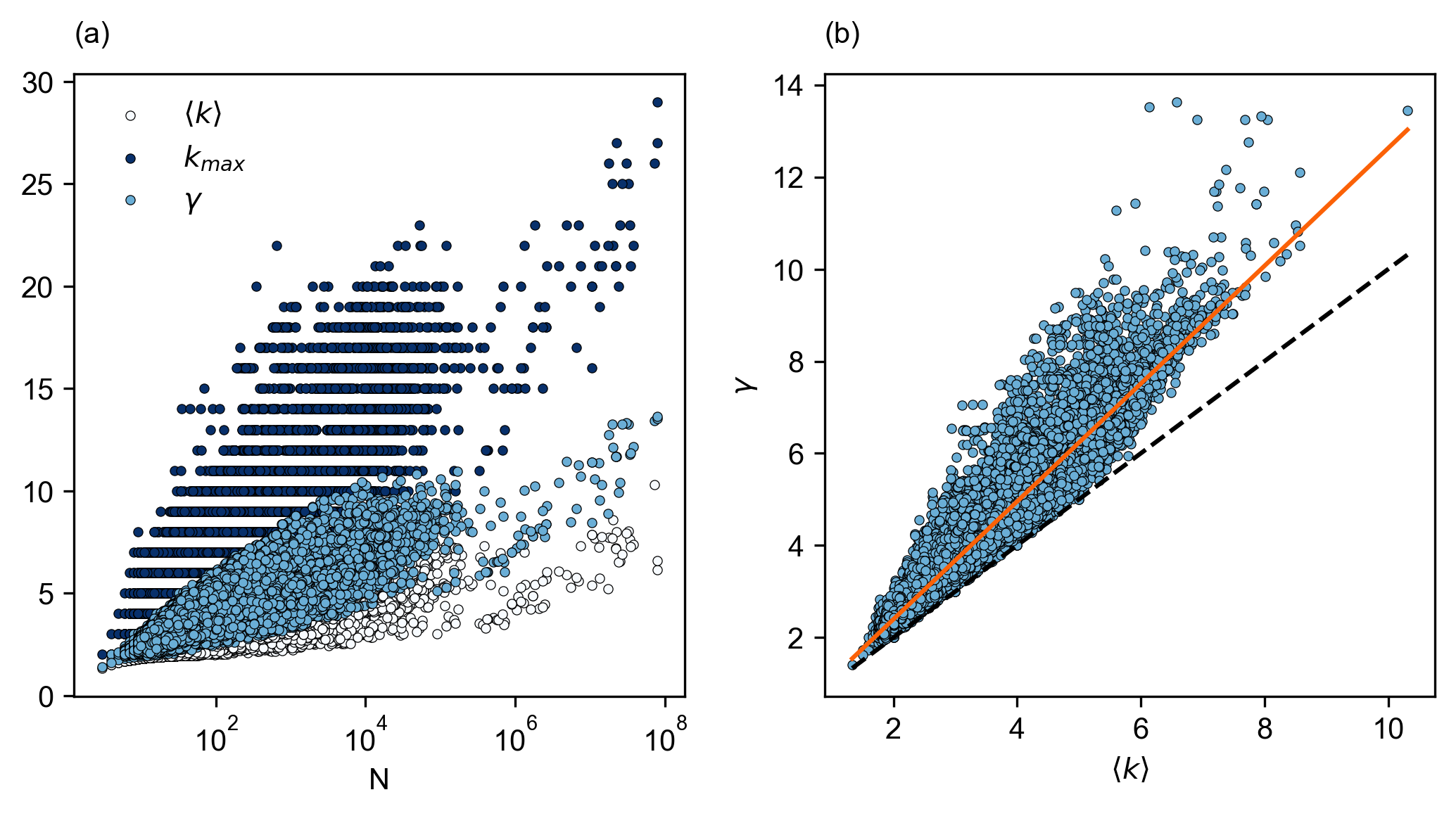}
    \caption{Topological quantities characterizing toyLIFE two-gene networks with a pattern-generating phenotype (T2P). (a) We represent the maximum and average network degree, as well as network spectral radius as a function of CC size, for CCs obtained through the exhaustive enumeration of the sequence space. (b) Connected component spectral radius $\gamma$ as a function of the corresponding average degree $\langle k \rangle$. The orange line represents a linear fit between the two measures: $\gamma = 1.28 \langle k \rangle-0.16$, with $R^2=0.87$. The dashed black line represents the line $\gamma=\langle k \rangle$, as a visual aid to confirm that $\gamma > \langle k \rangle$. 
    \label{fig:toyLIFEtopology}
}
\end{figure}

toyLIFE is a multilevel map from binary genomes, $A=2$, to Boolean gene regulatory networks (GRNs) \cite{arias:2014,catalan:2017tesis}. toyLIFE sequences code for genes that are translated into 2D compact proteins following the rules of a hydrophobic-polar (HP) model for protein folding~\cite{li:1996}. These proteins interact to form dimers and, jointly, they alter the expression of genes, thus yielding Boolean GRNs. Given a suitable environment, these emerging GRNs may further define a simple metabolism, or cellular automata that generate in turn a variety of spatio-temporal patterns \cite{catalan:2018,catalan:2020}. Therefore, phenotype in toyLIFE can be defined at several levels that start at GRNs.   
 
\begin{figure}[b!]
    \centering
    \includegraphics[width=150mm]{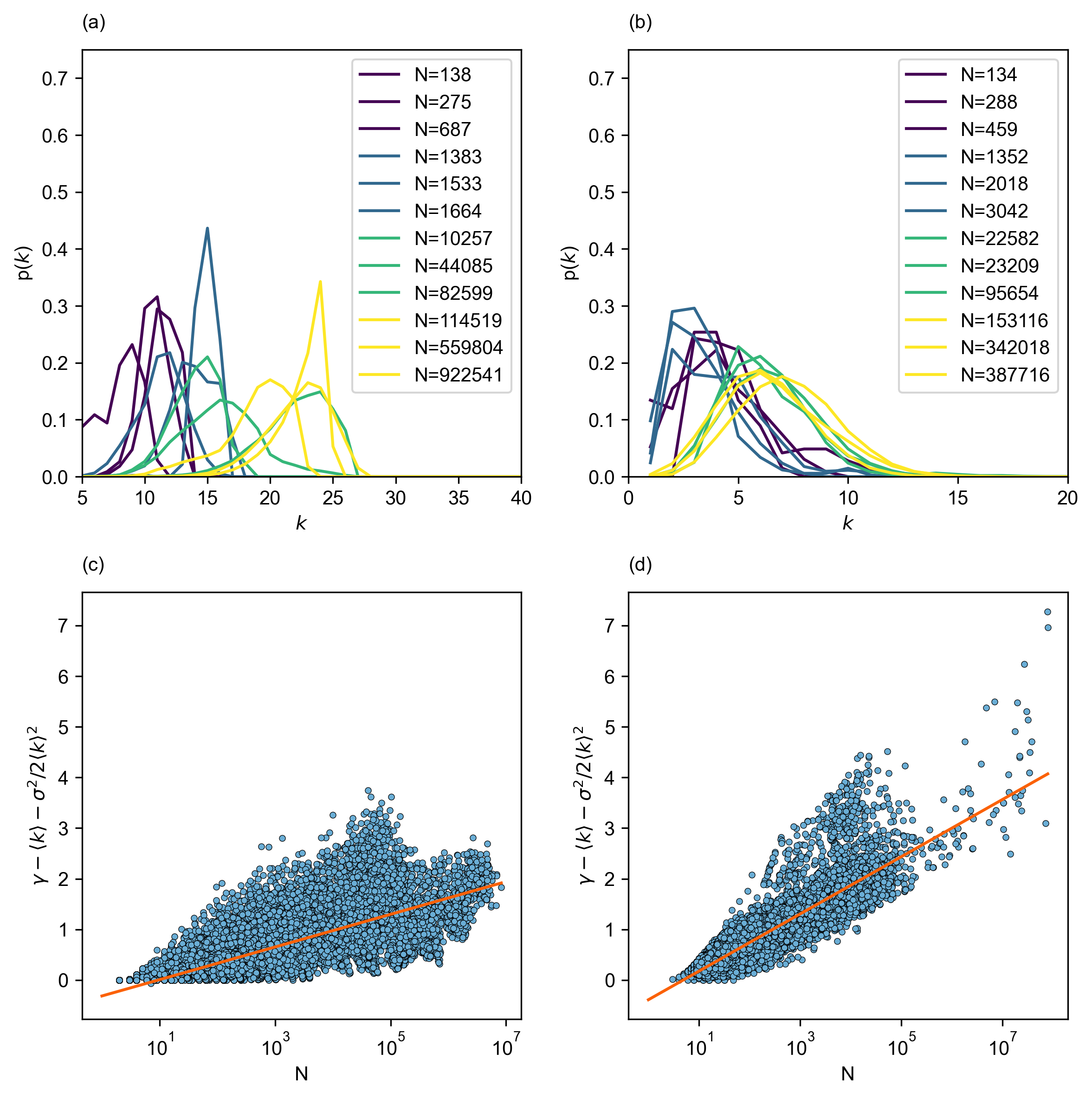}
    \caption{Degree distributions for RNA S3 and toyLIFE T2P and their effects on bounds to the spectral radius. (a) Degree distribution for RNA S3 with $L=16$; (b) degree distribution for toyLIFE T2P; difference $(\gamma-\langle k \rangle -\sigma^2/2\langle k \rangle^2$ for (c) RNA S3 with $L=16$ and (d) toyLIFE T2P. In (c) and (d), the orange line represents a linear fit: $\gamma-\langle k \rangle -\sigma^2/2\langle k \rangle^2= 0.32 \log_{10} N -0.32$ for RNA S3 ($R^2=0.56$) and $\gamma-\langle k \rangle -\sigma^2/2\langle k \rangle^2= 0.56 \log_{10} N -0.39$ for toyLIFE T2P ($R^2=0.79$).
    \label{fig:degree-sigma}
}
\end{figure}

In this work, we study two-genes toyLIFE with phenotype defined at the pattern-generating level (T2P) \cite{catalan:2020}. Each gene has a promoter region (4 positions) plus 16 positions coding for a protein (yielding $4 \times 4$ compact lattice proteins). Hence, $L=40$ and $k_{\rm max} \leq 40$. In T2P, toyLIFE cells are embedded in a 1D tissue where proteins can diffuse to neighboring cells. This gives rise to spatiotemporal expression patterns across the tissue, that constitute the phenotype. 

We have selected toyLIFE as a limit example of a complex, yet tractable, GP map that, in its two-gene version, might be severely affected by finite size effects. Still, the qualitative behavior of topological quantities of toyLIFE T2P phenotype networks is equivalent to that described for RNA S3, though the multilevel nature of toyLIFE T2P yields a larger dispersion in the relationships measured. This can be quantitatively observed in  Figure~\ref{fig:toyLIFEtopology}(a), which represents the values of $k_{\rm max}$, $\langle k \rangle$ and $\gamma$ for the 12,051,440 CC analyzed in toyLIFE T2P. Note that these are not all CC in T2P, only those obtained from phenotypes with $N<10^8$ for computational efficiency). Figure~\ref{fig:toyLIFEtopology}(b) depicts the relationship between the average degree and the spectral radius. This latter figure indicates that only relatively small CC are homogeneous, since there are no instances of large CC fulfilling the equality $\gamma = \langle k \rangle$. As expected, the observed dispersion is smaller in RNA S3, where all CC remain closer to the diagonal, as shown in Fig. \ref{fig:RNAtopology}(b).  

Finally, we represent in figure~\ref{fig:degree-sigma}a and b several degree distributions in the two GP maps studied to illustrate their variation with the specific phenotype for finite $L$, even in phenotypes of comparable size.  
The obtained distributions are relatively peaked around a well-defined average, so the correction obtained by including $\sigma$ is small. This is further illustrated in Figure~\ref{fig:degree-sigma}c and d, where the bound given by Eq.~(\ref{eq:sigmabound}) is depicted. There is no noticeable improvement with respect to the results reported in Fig. \ref{fig:RNAtopology-dif} when the dispersion of the degree distribution is included in the bound. 

\section{Numerical examples of phenotypic transitions}
\label{sec:numerical-dynamics}

 Estimating the eigenvalue $\lambda$ of genotype networks in fitness landscapes is difficult for at least two reasons. First, an exhaustive enumeration of all genotypes in a given phenotype is out of reach even for relatively short sequences; second, even if the replicative ability of genotypes is known, the calculation of the spectral radius of large networks is a costly computational procedure. The use of the log size of the phenotype as a proxy for the average degree $\langle k \rangle$, first, and then for $\gamma$ seems feasible in the light of the numerical results in the previous Section. Under these consecutive assumptions, Eq. \ref{eq:r-N} estimates the transition point between two phenotypes given their replicative abilities and their sizes. Though this estimation will be necessarily worse than that obtained through $\lambda$, it informs on the transition point in many situations where the full degree distribution of a genotype network is not available, or when the  coefficients involved in a relationship such as Eq. (\ref{eq:RNA-kN}) are unknown. Therefore, the advantage of using $\log N$ instead of $\gamma$ comes from the existence of various low-cost computational methods that allow accurate \cite{jorg:2008} and approximate \cite{garcia-martin:2018,martin:2022} estimations of phenotype size. 

We have explored transitions between phenotypes in RNA S3 and toyLIFE T2P in a two-peak landscape to characterize the transition and, chiefly, to quantify how the transition point depends on the topological characteristics of the phenotype. A schematic of the scenario studied is represented in Figure~\ref{fig:cartoon}. Our exhaustive enumeration of the two GP maps described (RNA S3 and toyLIFE T2P) allows to characterize all phenotype networks and the links between phenotypes, that is, nodes that belong to each of the phenotypes, and the complete set of neighboring phenotypes one mutation away ---with at least one connecting pair, but few to many in general. The transition matrix ${\bf M}_{\alpha+\beta} = (1-\mu) {\bf R}_{\alpha+\beta} + (\mu/S) {\bf G}_{\alpha+\beta} {\bf R}_{\alpha+\beta}$, with ${\bf G}_{\alpha+\beta}$ as depicted in Figure~\ref{fig:cartoon}(c); ${\bf M}_{\alpha+\beta}$ has dimension $(N_{\alpha}+N_{\beta})^2$ and $\lambda_{\alpha+\beta}$ is its largest eigenvalue, with associated eigenvector ${\bf v}=(v_i)$. 

The first approximation we made was that matrix ${\bf M}_{\alpha+\beta}$ would have a block-like structure, with few, off-diagonal, non-zero elements, following \cite{schuster:1988}. In an updated representation~\cite{aguirre:2013} the two-phenotype system has been described as two connected networks ``competing'' for ``resources'' (in the present case, resources correspond to replicators). It has been shown~\cite{aguirre:2013} that the eigenvector centrality of connector nodes, in our case the set of genotypes that are one mutation away but belong to different phenotypes, determines how sharp is the transition at $\lambda_{\alpha}=\lambda_{\beta}$. The larger the number of connector nodes and their eigenvector centrality, the less accurate becomes the prediction based on the two-block separation. 

\begin{figure}[b!]
    \centering
    \includegraphics[width=150mm]{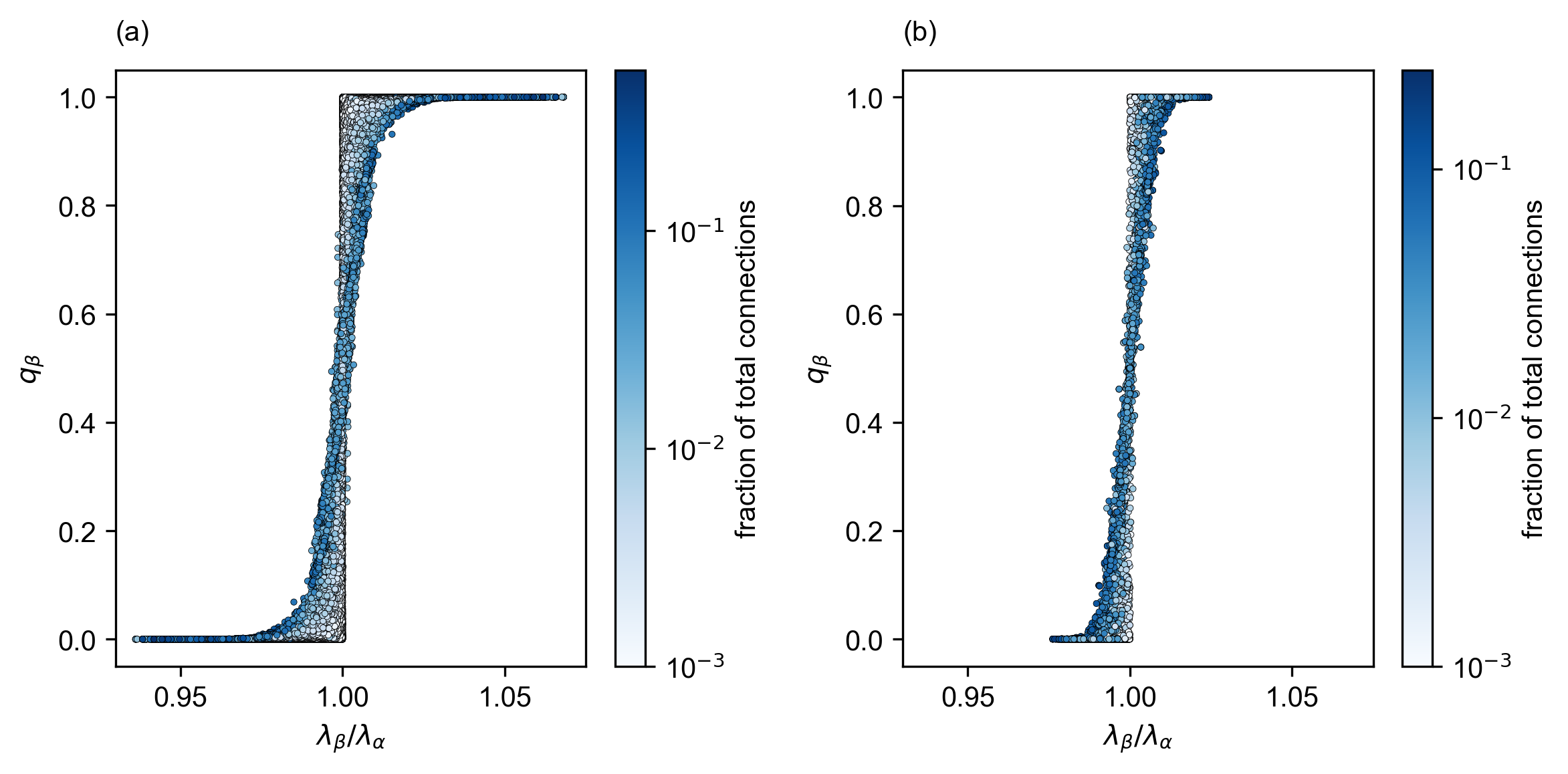}
    \caption{Accuracy of the prediction $\lambda_{\alpha}=\lambda_{\beta}$ as transition point. (a) RNA. Each point represents a system of two phenotypes of different size, each characterized through its fitness $\lambda$. The vertical axis shows $q_{\beta}$, the fraction of replicators in phenotype $\beta$. All 206,821 pairs of the 3,311 CCs of RNA S3 with $L=14$ that are mutually accessible through point mutations were used for this plot. (b) As previous panel, for toyLIFE. The color scale represents the fraction of links that connect every pair of phenotypes. Most pairs of phenotypes follow the prediction, with a dispersion of about $1\%$ around $\lambda_{\alpha}=\lambda_{\beta}$.}
    \label{fig:RNA-lambdaTransition}
\end{figure}

Let us define $q_{\alpha} = U^{-1} \sum_{i=1}^{N_{\alpha}} v_i$ as the fraction of the population of replicators that occupies nodes in phenotype $\alpha$ at equilibrium. Also, there is a fraction $q_{\beta} = U^{-1} \sum_{i=N_{\alpha}+1}^{N_{\alpha+\beta}} v_i = 1-q_{\alpha}$ of the population of replicators occupying nodes in phenotype $\beta$, with $U=\sum_{i=1}^{N_{\alpha+\beta}} v_i$. Figure \ref{fig:RNA-lambdaTransition} shows, for many different pairs $(\alpha,\beta)$ of phenotypes, the fraction $q_{\beta}$ at equilibrium. Most of the population occupies nodes in phenotype $\beta$ when $\lambda_{\beta}>\lambda_{\alpha}$, and {\it vice versa}. The color scale represents the fraction of links that actually connect both phenotypes in relation to its maximum possible number, showing that this fraction is in the majority of cases small. As expected, the transition around $\lambda_{\alpha}=\lambda_{\beta}$ is sharp for pairs of phenotypes weakly connected, while large fractions of connector links between phenotypes smoothen the transition \cite{aguirre:2013}.

\subsection{Transition as a function of average degree}

This is a second approximation where we assume that the average degree is a good approximation of the spectral radius, $\langle k \rangle \simeq \gamma$. In previous sections, we have seen that this is not always the case since, though degree distributions are peaked around well-defined average values, genotype networks are heterogeneous in degree, and their heterogeneity does not vanish with increasing phenotype size. What is more important, the average degree depends on each specific phenotype, as we have seen explicitly with RNA, and numerically with the two examples we have explored. An advantage of using the average degree to predict the transition point, however, is that $\langle k \rangle$ can be obtained through suitable sampling of nodes in a genotype network, and does not require an exhaustive knowledge of the network ---which is needed to calculate $\gamma$ or $\lambda$. 

\begin{figure}[b!]
    \centering
    \includegraphics[width=150mm]{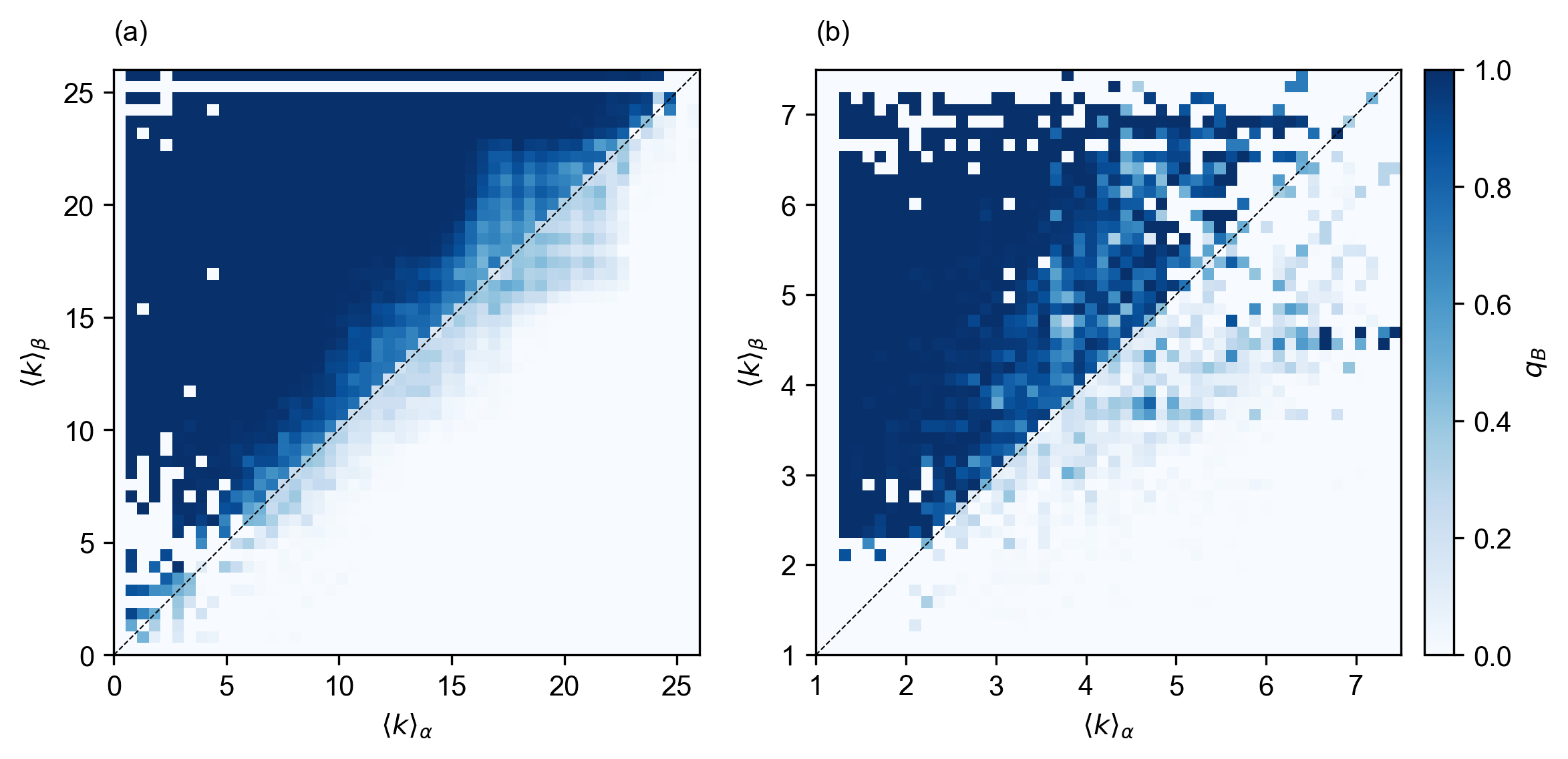}
    \caption{Accuracy of the prediction $\langle k \rangle_{\alpha}=\langle k \rangle_{\beta}$ as transition point (assuming $r_{\alpha} = r_{\beta}$). Results are represented in the plane $(\langle k \rangle_{\alpha},\langle k \rangle_{\beta})$. (a) RNA S3, $L=14$. Each point stands for an average over all pairs of phenotypes with the corresponding average degrees. The color scale indicates the fraction $q_{\beta}$ of the total population occupying nodes of phenotype $\beta$, averaged over pairs. (b) toyLIFE T2P. As in the previous plot. 
        \label{fig:RNA-degreeTransition}
}
\end{figure}

Figure \ref{fig:RNA-degreeTransition} represents the fraction $q_{\beta}$ averaged over all pairs of phenotypes with degree $(\langle k \rangle_{\alpha}, \langle k \rangle_{\beta})$. White points stand for non-existing pairs; statistics are better for average values of the degree, between 5 and 20 for RNA S3, $L=14$ and 3 and 5 for toyLIFE T2P. The prediction worsens close to the diagonal $\langle k \rangle_{\alpha} = \langle k \rangle_{\beta}$, though it is quite good for RNA and slightly worse for toyLIFE. We do not observe any improvement in the predicted transition point with larger average degree. The prediction is very good when the difference between the average degree of the two phenotypes is about 2-3 or larger. Despite all the caveats, the average degree yields a reasonable, probabilistic estimate, of the position of the transition between two phenotypes. 

\subsection{Transition as a function of phenotype size}

\begin{figure}[b!]
    \centering
    \includegraphics[width=150mm]{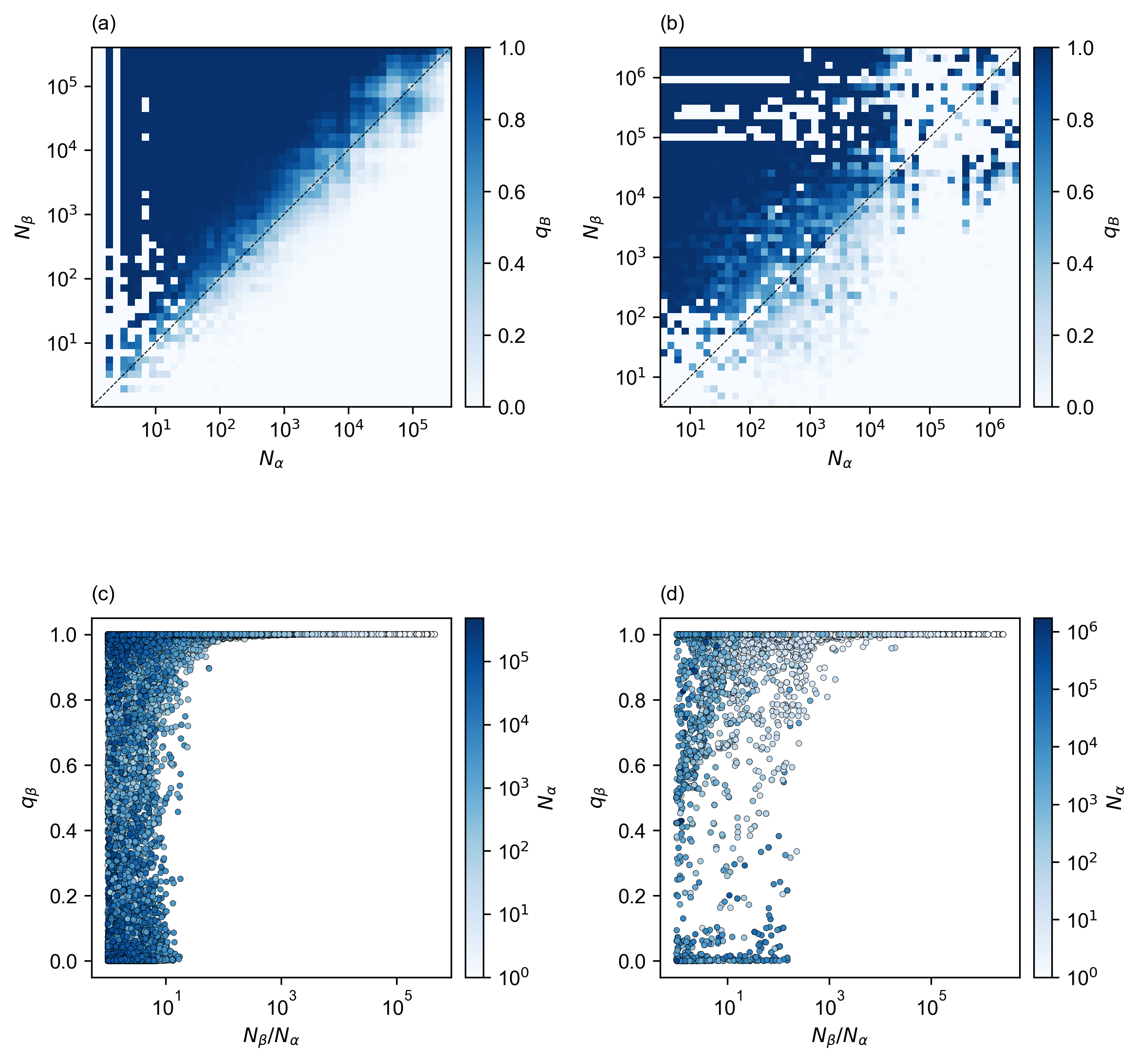}
    \caption{Accuracy of the prediction $\log N_{\alpha}=\log N_{\beta}$ as transition point (assuming $r_{\alpha} = r_{\beta}$ and a phenotype independent coefficient $c$). (a,b) Results are represented in the plane $(\log N_{\alpha}, \log N_{\beta})$ for (a) RNA S3, $L=14$ and (b) toyLIFE T2P. Each point stands for an average over all pairs of phenotypes with the corresponding sizes. The color scale indicates the fraction $q_{\beta}$ of the total population occupying nodes of phenotype $\beta$, averaged over pairs. (c,d) Fraction of population in phenotype $\beta$, $q_{\beta}$ as a function of the relative size between the two phenotypes, $N_{\beta}/N_{\alpha}$. The color scale indicates the size of the smallest phenotype in the compared pair. (c) RNA S3, $L=14$; (d) toyLIFE T2P. 
    \label{fig:RNA-sizeTransition}
}
\end{figure}

 The third and last approximation we check here is the substitution of the average degree of a network by the logarithm of the phenotype size, times a phenotype-dependent multiplicative term, $\langle k \rangle \simeq c \log N$ to predict the transition point. As discussed above, there are reasons to assume that $c$ becomes asymptotically independent of the phenotype for typical phenotypes, at least in RNA. We cannot discard that some GP maps may behave otherwise though, as of yet, we do not have examples contradicting that assumption. Figure \ref{fig:RNA-sizeTransition}a, b represents the fraction $q_{\beta}$ averaged over all pairs of phenotypes with sizes $(\log N_{\alpha}, \log N_{\beta})$, while Figure \ref{fig:RNA-sizeTransition}c, d represents individual pairs. This prediction could be improved had we included the value of the coefficient $c$ for different phenotypes; however, we have chosen to represent the case where $c$ is assumed to be phenotype-independent as a limit case with the minimum number of quantities to estimate: just phenotype size. 

The use of $N$ as a proxy to estimate transitions between phenotypes has a practical and a conceptual implication, as we have anticipated in previous sections. On the practical side, phenotype size can be easily estimated with a variety of methods of different accuracy available in the literature \cite{jorg:2008,garcia-martin:2018,martin:2022}; on the conceptual side, the relationship between the transition point and phenotype size provides a quantitative measure of the importance of phenotype redundancy, a measure of entropy, in phenotype fitness. Though this prediction is not as good as the one obtained if the whole genotype network is known, it is reasonable attending to the computational effort needed to estimate phenotype size. Further, it allows a first estimation of the relative adaptive value of replicative ability versus phenotype size. For RNA S3 and $L=14$, an order of magnitude difference in phenotype size means that the smaller phenotype is essentially empty; whether this difference remains constant or increases with $L$ remains to be explored. For toyLIFE T2P, the difference required is slightly larger, about 1.5 orders of magnitude. For pairs of phenotypes more similar in size, however, the larger one typically attracts over $50\%$ of the population ---note that light-blue points are rare in any case. 

\section{On finding a sufficiently fit phenotype}

Evolution is severely conditioned by the size of phenotypes, as all studies with synthetic and empirical GP maps have demonstrated. In the sections above, we have derived quantitative relationships between the fitness $\lambda$ and two important features: replicative ability and phenotype size. The dependence with phenotype size $N$ provides a first explanation of why large phenotypes are the only ones seen by natural selection: when a population has to choose between two phenotypes of comparable replicative ability, the larger one will be preferred. The question arises: how much replicative ability is lost as a consequence of phenotype size? 

It is another quasi-universal property of GP maps that explains the huge span in phenotype sizes: the distribution of phenotype sizes is in most cases well fit by a lognormal function~\cite{dingle:2015,manrubia:2017,catalan:2017,garcia-martin:2018}. This feature of phenotype sizes entails that there are orders of magnitude difference between abundant, typical, and rare phenotypes, even for relatively short sequences: an astronomically large number of phenotypes invisible to evolution. The preference for large phenotypes can be dramatically illustrated with the case of natural, non-coding RNA sequences \cite{dingle:2015}. For example, most abundant phenotypes in sequences of length $L=126$ have sizes between $10^{20}$ and $10^{40}$. However, phenotypes selected in natural systems (meaning here RNA S3 sequences available at the fRNAdb \cite{kin:2007}) have sizes not smaller than $10^{36}$, reaching $10^{46}$ in many cases \cite{dingle:2015}. For other functional, non-coding RNAs subjected to strong selective pressures on the secondary structure, such as viroids \cite{diener:1971,flores:2012}, phenotype sizes can reach $10^{90}$ for $L \simeq 399$ \cite{catalan:2019a}. The number of compatible genotypes rapidly becomes hyperastronomically large for any realistic functional phenotype \cite{louis:2016}.     

\subsection{Visible values of phenotype size}

\begin{figure}[b!]
    \centering
    \includegraphics[width=130mm]{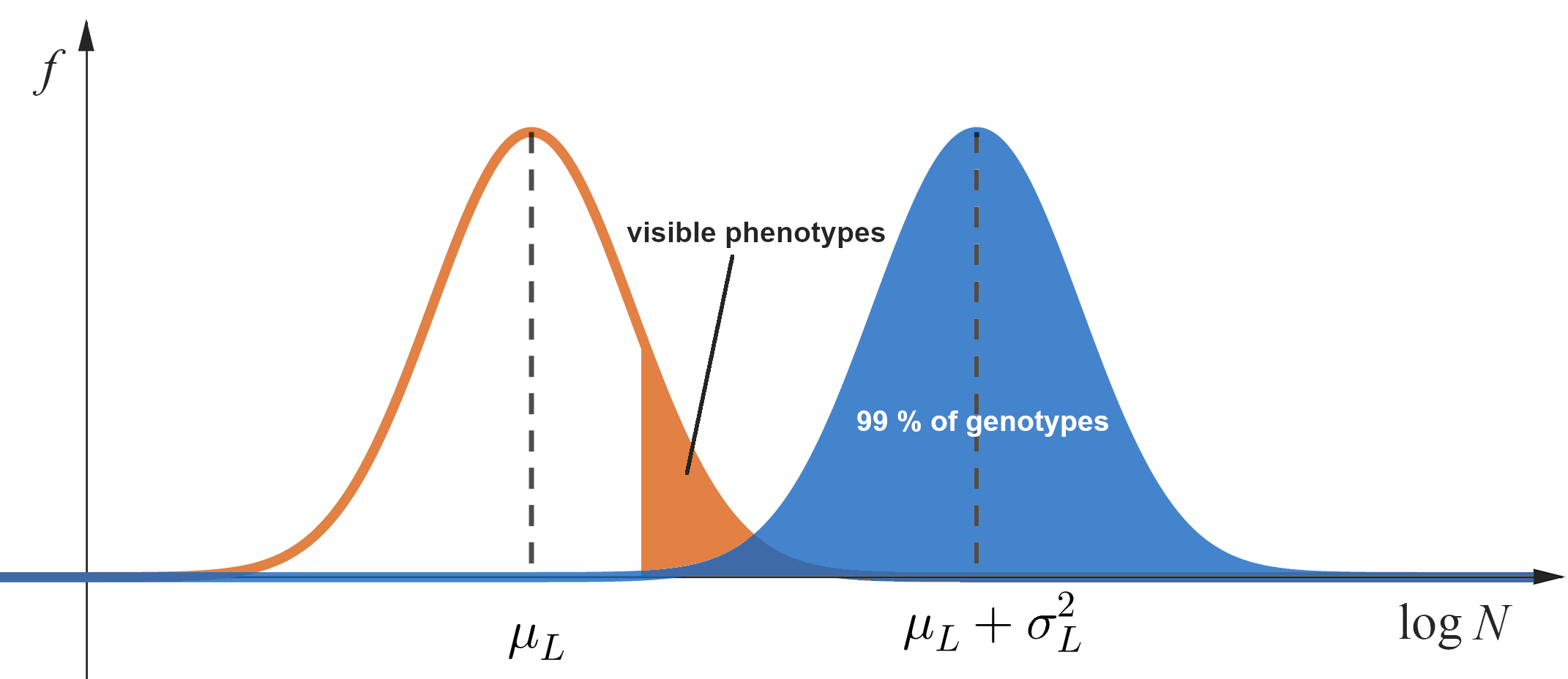}
    \caption{Fraction of phenotypes $f$ with a log-size $\log N$ (orange), as well as the fraction of genotypes in phenotypes of log-size $\log N$ (blue), for sequences of length $L$. The former are normally distributed with mean $\mu_L$ and standard deviation $\sigma_L$; the latter are distributed as $Nf(\log N)$---hence as a normal of mean $\mu_L+\sigma_L^2$ and standard deviation $\sigma_L$ \cite{garcia-martin:2018}. The ``visible'' phenotypes are those in which nearly all genotypes (here $99$ \%) are concentrated. 
    }
    \label{fig:gaussians}
\end{figure}

Although the following discussion applies to any GP map with a log-normal distribution of phenotype size, 
for the sake of illustration ---and because it is the best documented example in the literature---, we will focus on the case of the RNA S3 map. For this map we know \cite{garcia-martin:2018} that the fraction of phenotypes with a given $\log N$ is a normal distribution with mean $\mu_L$ and standard deviation $\sigma_L$, given by
\begin{equation*}
    \mu_L=\mu_1L+O(1), \qquad
    \sigma_L=\sigma_1L^{1/2}+O\left(L^{-1/2}\right),
\end{equation*}
where $\mu_1=0.2865$ and $\sigma_1=0.4434$. This means that the fraction of genotypes belonging to a phenotype of log-size $\log N$ is also a normal distribution with the same variance but shifted up to a mean value $\mu_L+\sigma_L^2$ \cite{garcia-martin:2018}. The two distributions are sketched in figure~\ref{fig:gaussians}. This picture illustrates why most genotypes (say 99\% or 99.9\% of them) are only found in a fraction of the largest phenotypes (the orange-colored region in the figure). In particular, it explains why real phenotypes found in nature belong to this top region of the size distribution \cite{dingle:2015}.

We can make this argument more quantitative. Let $p$ be the fraction of genotypes in the blue region of figure~\ref{fig:gaussians} (as we said, $p=0.99$ or $p=0.999$). If $N_p$ is the lowest size limiting this region from below, then
\begin{equation}
    \log N_p=\mu_L+\sigma_L^2-\sqrt{2}\sigma_Lz_p, \qquad p=1-\frac{1}{2}\erfc(z_p),
\end{equation}
$\erfc(z)$ being the complementary error function. From this size we can now compute the fraction of phenotypes to which those genotypes belong (the orange region of figure~\ref{fig:gaussians}) as
\begin{equation}
    q=\frac{1}{2}\erfc(\zeta_p), \qquad \zeta_p=\frac{\log N_p-\mu_L}{\sqrt{2}\sigma_L}
    =\frac{\sigma_L}{\sqrt{2}}-z_p.
\end{equation}
This is the fraction of ``visible'' phenotypes---those that evolution can find in a random exploration of the genotype space. Table~\ref{tab:zps} shows that $z_p$ is not very sensitive to the precise value of $p$, so a reasonable estimate for it is $z_p\approx 2$.

\begin{table}[t]
    \centering
    \begin{tabular}{ll}
        \hline
        \hline
        \multicolumn{1}{c}{$p$} & \multicolumn{1}{c}{$z_p$} \\
        \hline \\[-4mm]
        0.99 & 1.644976 \\
        0.999 & 2.185124 \\
        0.9999 & 2.6297417 \\
        \hline
    \end{tabular}
    \caption{Values of $z_p$ for different choices of the fraction $p$.}
    \label{tab:zps}
\end{table}

Now, an asymptotic approximation of $\erfc(u)$ when $u\to\infty$ is \cite[eq.~7.12.1]{NIST:DLMF}
\begin{equation*}
    \erfc(u)\sim\frac{e^{-u^2}}{\sqrt{\pi}u}\big[1+O(u^{-2})\big],
\end{equation*}
therefore, as $L\to\infty$,
\begin{equation*}
    q\sim\frac{1}{2\sqrt{\pi}}\frac{\sqrt{2}}{\sigma_L-\sqrt{2}z_p}
    e^{-\left(\frac{\sigma_L}{\sqrt{2}}-z_p\right)^2}
    \sim\frac{1}{\sqrt{2\pi}\sigma_1L^{1/2}}e^{-\sigma_1^2L/2}
    =\frac{0.9}{L^{1/2}}(1.1)^{-L}.
\end{equation*}
In other words, the fraction of visible phenotypes for a fixed, large value of the fraction of genotypes, \emph{decreases exponentially} with the length $L$ of the genotype. (Notice how this asymptotic estimate \emph{does not} depend on the actual value of $z_p$.)

But, on the other hand, the total number of phenotypes for RNA secondary structures (with stacks formed by at least two consecutive nucleotide pairs and terminal loops with at least three unpaired nucleotides) grows with $L$ as $1.48L^{-3/2}(1.85)^L$~\cite{stein:1979,schuster:1994,cuesta:2017}. Therefore, the absolute number of different, large phenotypes covered by a fraction $p$ of genotypes actually \emph{grows exponentially} with $L$ as $1.33L^{-2}(1.68)^L$.

In summary, the visible phenotypes are only a negligible fraction of all the possible phenotypes that could potentially exist, and nevertheless, the absolute number of them is still huge. So evolution has a lot of variability to choose from even if it only ``sees'' a tiny bit of it. But the question remains whether a high replicative ability can compensate for this ``blindness'' so as to bring any of these hidden phenotypes to light. In the next subsection we will explain why we think this is highly unlikely to happen.

\subsection{Attainable values of phenotype replicative ability}

The distribution of replicative abilities of possible phenotypes is mostly unknown, but its range of values can be guessed based on empirical evidence. A paradigmatic example of increase in replicative ability is provided by Spiegelman's experiment where, allowing for arbitrary changes in their length, RNA sequences attained a 15-fold increase in replicative speed \cite{mills:1967}. In a more realistic cellular environment, a measure of replicative ability is given by the processivity of RNA Pol II. This protein synthesizes RNA at a speed between 1 kb/min and 6 kb/min, with a clear peak around 3 kb/min and little difference between genes \cite{muniz:2021}. It seems reasonable to assume that biochemical constraints bound the possible values of the replicative ability of phenotypes, even if other traits are under selection, to a relatively narrow range that spans from a few-fold increase to an order of magnitude. 

For the sake of simplicity, let us therefore assume that the replicative ability of a set of phenotypes (for example, those able to accomplish a specific task) follows a Gaussian distribution with average $\langle r \rangle$ and variance $\sigma_r^2$. Then, the maximum value of $M$ occurrences follows a peaked distribution around the average value $h_M\sim r+\sigma_r\sqrt{2\log M}$ \cite{haan:2006}. If $M \sim b L^{\alpha} a^L$, then $h_M \sim r+\sigma_r\sqrt{2 L \log a}$. Let us now compare the average value of two sets, the first one including all possible phenotypes $M \sim 1.48 L^{-3/2} (1.85)^L$, and the second one embracing those phenotypes covering nearly all genotypes, $M' \sim 1.33 L^{-2} (1.68)^L$,  as calculated in the previous section. Thus, $h_M \sim r+1.11\sigma_rL^{1/2}$ and $h_{M'} \sim r+1.02\sigma_rL^{1/2}$. The latter is not even 10\% lower than the former. In other words, among the accessible phenotypes the population can find phenotypes with replicative abilities comparable to the largest ones available in the whole phenotype space. 

\section{Discussion and conclusions}

Phenotypes can be described as connected networks of genotypes mutually accessible through mutations. In fixed environments, the fitness of a phenotype corresponds to the largest eigenvalue of the transition matrix associated to the network of genotypes. In this contribution we have shown that, in the simplified case where all genotypes in a phenotype have the same replicative ability, the transition between two phenotypes can be successively approximated, with decreasing precision, by the relationship between the two eigenvalues of the phenotypes, the average degree of their genotype networks and finally the log-size of the phenotypes. This latter case is interesting due to the existence of simple computational methods to estimate the size of a phenotype and, especially, because it measures the quantitative relevance of phenotype size in adaptation. In the current context, phenotype size is a measure of entropy and also of robustness of the phenotype \cite{dingle:2015} and, as such, its turns out to be an essential component of phenotype fitness. Updated representations of fitness landscapes that include the networked nature of phenotypes---such as adaptive multiscapes \cite{catalan:2017}---become essential to re-educate our intuition on the outcomes of the evolutionary process. 

Our approach to the description of the transition between phenotypes has been necessarily simple. We have considered a two-peak landscape for replicative ability and calculated the eigenvector of the joint transition matrix, which represents mutation-selection equilibrium. There are multiple studies that, inspired by the overarching concept of punctuated equilibria \cite{eldredge:1972}, have explored the speed of the transition of a population of replicators between two loosely connected networks. In all such studies, sudden transitions in genotype spaces have been identified~\cite{aguirre:2018}. Early descriptions of sudden transitions corresponded to adaptation to increasingly fitter phenotypes~\cite{schuster:1988,huynen:1996}, akin to the scenario explored here. In neutral networks with community structure, the population mostly concentrates in the largest community~\cite{capitan:2015}, though sudden transitions occur every time a larger community is found~\cite{wilke:2001BMB}. This phenomenology is analogous to that observed in rough fitness landscapes with complex topology, where the largest fraction of the population is found within a small subset of connected nodes, experiencing sudden shifts in genome space under smooth environmental changes \cite{aguirre:2009}, even if phenotypes are not explicitly defined~\cite{aguirre:2015}. Numerical simulations of the transition between two phenotypes when the replication rate is smoothly varied (results not shown) yield fast transitions of the type described in previous works. All these observations are well understood in a theoretical framework where networks are visualized as ensembles that compete for resources \cite{aguirre:2013}. Transitions between two such networks can be smooth or sudden, with all possibilities in between, depending on the number of connector links between the networks and the eigenvector centrality of the connecting nodes. Therefore, the strength of the transition and the fraction of population in either network can be tuned through an appropriate election of the nodes than link one network to another. The fact that adaptive transitions in populations of replicators embedded within a GP map are sudden is consistent with the existence of a reduced number of links between phenotypes and the expectation that most of these connections link peripheral genotypes. The topology of genotype networks, unlike in synthetic examples of neutral networks \cite{wilke:2001BMB}, cannot be modified at will. Community structure seems to be a generic property of realistic GP maps, be these communities mutually neutral, different phenotypes, or a subset of nodes in a fitness landscape. If this is so, sudden transitions in genotype spaces should be the rule also in natural systems (see e.g.~\cite{koelle:2006})---though it might be difficult to disentangle the role played by various variables, such as phenotype size, replicative ability, environmental changes (which may modify at once the two previous variables~\cite{catalan:2017}), or increases in robustness. 

Large phenotypes embrace various evolutionary advantages, not all of them adaptive. First, there is a dynamical advantage, known as phenotypic bias, due to the fact that the typical discovery time of a phenotype in blind searches is proportional to the inverse of its frequency \cite{schaper:2014}. Recent studies have related this phenotype bias to a simplicity bias, arguing that phenotypes with many genotypes (resulting from a GP map) have to be simple in terms of algorithmic information theory and Kolmogorov complexity \cite{dingle:2018,johnston:2022}. This interesting relationship provides an additional way of predicting transitions between phenotypes, where phenotype size would be substituted by phenotype complexity, a quantity also simple to estimate from a computational viewpoint \cite{dingle:2022JRSI}. Second, the average robustness of phenotypes (their average degree) increases with its size as $\langle k \rangle \propto \log N$; that is, the larger the phenotype, the more robust its nodes are. Higher robustness (higher entropy) confers an immediate adaptive advantage. Third, larger phenotypes also have further access to evolutionary novelty, by guaranteeing navigability of the genotype space and facilitating contact with a higher diversity of phenotypes. Altogether, selection of larger (hence fitter) phenotypes appears as an evolutionary trend that could entail a form of irreversibility in evolution. This process is also related to the unfathomable size of genotype spaces: networks of genotypes become so large, even for relatively short sequence lengths, that natural populations are unable to explore any significant portion of them, even in substantial evolutionary time, causing a perpetual drift to more robust regions: they are never stably sitting at the top of a hill. 

It has not escaped our notice that the consistent observation that only large phenotypes are found in natural RNA sequences (and probably in any realistic GP map) immediately suggests an alternative interpretation in the light of our results, namely, that the number of phenotypes with significantly larger replicative ability and typical (or smaller) size is actually negligible. In other words, the minimum size of phenotypes found in nature may actually bound the loss in replicative ability, and not the other way round. Should it be otherwise, why would smaller phenotypes with larger $\lambda$ not be gradually fixed through natural selection? In favor of this alternative view comes the observation of how powerful natural selection is to select for phenotypes that would be never found under blind searches \cite{dawkins:1996}, but that can be attained under parsimonious incorporation of increasingly rare (at least at first sight) solutions. Finally, it cannot be discarded that selection for improved replicative ability (or for optimized functionality) and selection for higher robustness (or higher phenotype size) occur concomitantly. If a phenotype highly optimized for function is too rare to guarantee sufficient robustness, the size of such phenotype could be enlarged through modification of traits that preserve function and increase size (though these are usually not included in simple models), such as genotype length \cite{cuypers:2012,cuypers:2014}, the emergence of additional levels in the GP map \cite{catalan:2018} or the formation of complex interacting molecular ensembles \cite{colizzi:2014}. Nature always finds a way.  

\ack
The authors acknowledge discussions with L. F. Seoane. This study was funded by grants PID2020-113284GB-C21 (S.M.), PGC2018-098186-B-I00 (P.C. and J.A.C.), PID2019-109320GB-100 (P.C.), and PID2021-122936NB-I00 (J.A.) all funded by MCIN/AEI/10.13039/501100011033 and by “ERDF A way of making Europe”.




\section*{References} 

\bibliography{bibliography}
\bibliographystyle{iopart-num}

\end{document}